\documentclass[twocolumn,trackchanges]{aastex631}
\usepackage{float}

\begin{document}

\title{Azimuthal Misalignments in Stellar Warp Structure as Dynamical Tracers of Mergers in Milky~Way--like Galaxies}

\author{Lekshmi Thulasidharan}
\affiliation{Department of Physics, University of Wisconsin--Madison, USA\\}
\author{Elena D'Onghia}
\affiliation{Department of Physics, University of Wisconsin--Madison, USA\\}
\affiliation{Department of Astronomy, University of Wisconsin--Madison, USA\\}
\affiliation{INAF--Osservatorio Astrofisico di Torino, Via Osservatorio 20, 10025 Pino Torinese (TO), Italy\\}
\author{Robert Benjamin}
\affiliation{Department of Astronomy, University of Wisconsin--Madison, USA\\}
\affiliation{Department of Physics, University of Wisconsin--Whitewater, WI, USA}

\begin{abstract}
We investigate the origin of warps in stellar disks using high-resolution Milky Way analogs from the IllustrisTNG50 simulation. Focusing on galaxies that experienced a major merger, we identify a characteristic azimuthal misalignment between the warp structures of stellar populations formed before and after the merger. This misalignment persists even after correcting for differential rotation, suggesting it is a dynamical imprint of the merger rather than a consequence of internal kinematics. In contrast, galaxies without significant merger events show no such offset between stellar populations of different ages. These findings support the scenario in which mergers can induce long-lived warps and leave detectable structural signatures in stellar disks. Applied to the Milky Way, this approach offers a potential way to test whether the Gaia-Sausage-Enceladus merger contributed to the formation of the Galactic warp. It may also provide an independent means to constrain the timing of such merger events by examining the phase offsets in the stellar warp as a function of stellar age.

 
\end{abstract}



\section{Introduction} \label{sec:intro}
\begin{figure*}[t!]
\centering
\includegraphics[width=1.03\textwidth ]{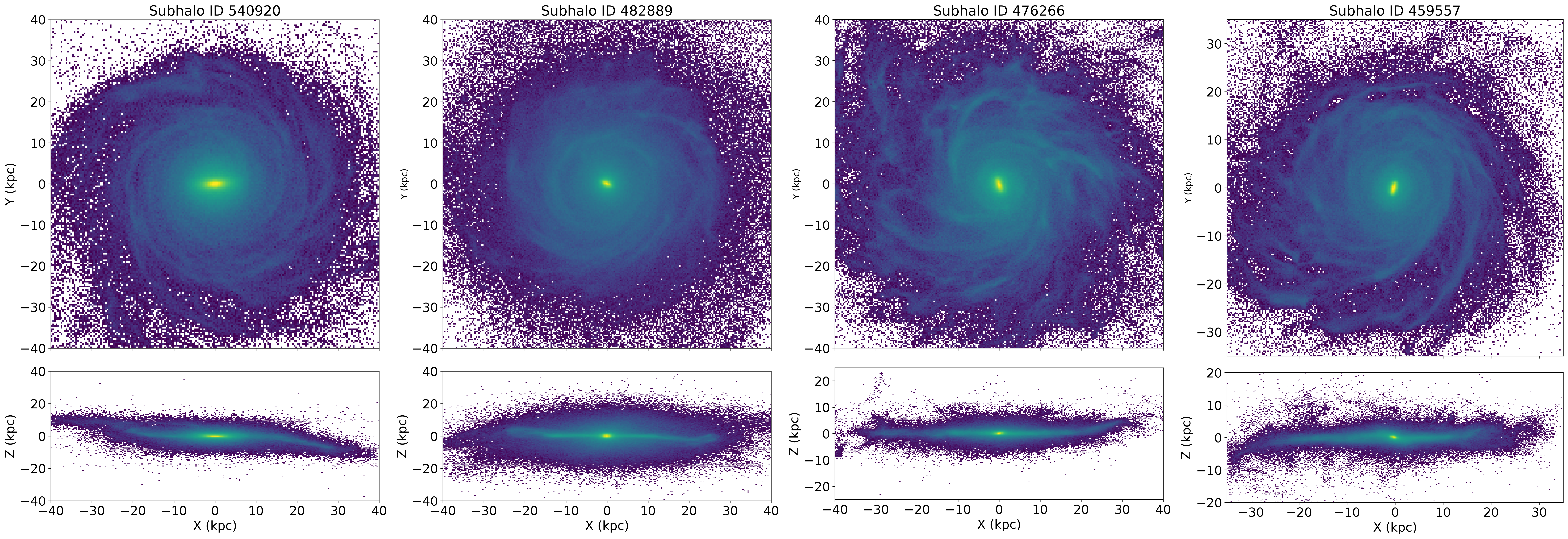}
\caption{Density projection of warped Milky Way analogs from IllustrisTNG50 in the X-Y and X-Z plane.}
\label{fig:fig1}
\end{figure*}
\par
Galactic disks are rarely perfectly flat. Instead, their outer regions often bend, forming warps, a ubiquitous feature in disk galaxies, including the Milky Way. The presence of a warp in the Milky Way has been established through HI surveys \citep{burke1957,Westerhout1957,kerr1957,kerr1957_b,levine2006} and stellar observations \citep{Bosma1978,Shostak1984,begeman1987,drimmel2000,alard2000,drimmel2001,lopez2002_a,lopez2014,robin2008,reyle2009,amores2017,poggio2018,huang2018,schronich2018,wang2018,skow2019,chen2019,gomez2019,chengdong2019,lopez2020,li2020,wang2020,yu2021,lemasle2022,nagi2022,ardevol2023,li2023,uppal2024,weixiang2025}, particularly with the advent of high-precision data from sky surveys such as Gaia \citep{gaia2016,gaia,gaia2023}. Additional confirmation comes from molecular gas studies \citep{grabelsky1987}, as well as interstellar dust maps \citep{freu1994}. Recent studies suggest that the Galactic warp is not static, but is precessing at 10--13 km s$^{-1}$ kpc$^{-1}$\citep{poggio2020,cheng2020,walter2023,jonsson2024}. Despite this, the physical origin of the warp remains a subject of ongoing debate. 

Several mechanisms have been proposed to explain its formation, including tidal interactions with satellite galaxies \citep{kim2014,gomez2016,poggio2020,semczuk2020}, such as the Large Magellanic Cloud \citep{Weinberg1995,weinberg1998,weinberg2006} and Sagittarius \citep{bailin2003b,laporte2018,binney2024}, misalignment between the galactic disk and a non-spherical dark matter halo \citep{ostriker1989,debattista1999,bailin2003}, continuous gas accretion from the intergalactic medium \citep{kahn1959,lopez2002}, the influence of the intergalactic magnetic fields \citep{battaner1998} and residual effects from past mergers \citep{Han_2023,han2023b,deng2024}. Finally, time‑dependent torques from a rotating or slowly tumbling, triaxial dark‑matter halo, which is common in cosmological simulations, can also generate azimuthally asymmetric warps by inducing a slowly precessing line of nodes \citep{ostriker1989,debattista1999,bailin2003}.

Among these scenarios, mergers present a compelling explanation, particularly given the evidence that the Milky Way experienced a major accretion event early in its history. The Gaia-Sausage-Enceladus (GSE) merger, which occurred 8--10 billion years ago, has been shown to have significantly influenced the Milky Way’s thick disk and inner halo structure \citep{helmi,bel}.

Beyond the Milky Way, warps have been observed in external galaxies, often in both their gaseous and stellar components. HI and optical surveys reveal a wide range of warp morphologies in disk galaxies \citep{Sancisi1976,Sancisi1983,briggs1990,resh1998,garcia2002,gentile2003,Ann2006,jozsa2007,resh2016,resh2025}, with some systems exhibiting long-lived distortions suggestive of sustained dynamical perturbations. Many edge-on spiral galaxies, such as NGC 4013 \citep{bottema1987}, show clear warp signatures in their stellar and gaseous components. However, the extent to which these features are linked to past mergers remains uncertain.

If mergers contribute to warp formation, they should leave a kinematic imprint in the form of an azimuthal misalignment: a shift in the orientation of the warp between stellar populations formed before and after the merger. This effect arises because the merger perturbs the angular momentum of the host galaxy, reorienting the disk and creating a measurable difference in the warp structure of different stellar populations. Quantifying this misalignment provides a potential dynamical record of past mergers, offering a direct observational test of the connection between warps and accretion events. Theoretical studies and simulations provide a means to test whether major mergers naturally induce azimuthal misalignments in stellar warps and whether such features persist over cosmic time.

To explore this question, we analyze Milky Way analogs from the high-resolution IllustrisTNG50 simulations, focusing on galaxies that have undergone major mergers. By examining the structure of their stellar warps, we assess whether post-merger populations exhibit systematic azimuthal shifts relative to pre-merger stars. We measure the magnitude of this misalignment and compare the results to those of a galaxy that has not undergone a major merger. Through this approach, we aim to establish whether azimuthal misalignment is a dynamical tracer of merger events and whether similar signatures could be detected in future observational studies of the Milky Way and external galaxies.

\section{Milky Way Analogues in IllustrisTNG50} \label{sec:floats}
Cosmological simulations provide key insight into how various galactic parameters evolve and respond to external disturbances, making them essential tools for understanding the dynamics and structure of the galaxy. IllustrisTNG50, the latest in the IllustrisTNG suite, combines high spatial resolution within a 50 Mpc box sampled by 2160$^3$ gas cells with a baryon mass resolution of $8.5 \times 10^4 M_{\odot}$ \citep{nels1,nelson2019,Pil}. Given its spatial resolution on par with the modern zoom-in simulations, TNG50 is a valuable resource for such studies. For our analysis, we used the publicly available catalog of Milky Way and Andromeda analogs from the IllustrisTNG50 simulation \citep{Pillepich2024}.

\subsection{Models}

Our analysis considers four warped model galaxies analogous to the Milky Way with Subhalo IDs 540920, 482889, 476266, and 459557, having total masses of $1.19\times10^{12} M_\odot$, $1.89 \times 10^{12} M_\odot$, $1.10 \times 10^{12} M_\odot$, and $0.8743 \times 10^{12} M_\odot$, respectively. The X-Y and X-Z density projections of the present-day snapshots for these models are shown in Figure \ref{fig:fig1}. The first three galaxy models underwent a major merger interaction, resulting in the formation of a warp. We chose the three models such that the galaxy doesn't undergo any other significant interactions to ensure they evolve secularly post-merger event, since we are interested in the warps induced by a major merger. While TNG50 contains many Milky Way--like galaxies, a systematic search for all such warped galaxies with an isolated merger history is computationally intensive. Therefore, we focused on a small subset of model galaxies that satisfied our constraints and displayed clear warp signatures. Model ID 459557 also has a warp; however, it seems to have formed due to reasons other than a major merger event, as it hasn't undergone one in its entire history based on its merger tree. We use it as our control model. 

Subhalo ID 540920 experiences a major merger event with a mass ratio of approximately 2.5:1, occurring at a lookback time of around 4 billion years. Similarly, Subhalo ID 482889 undergoes a merger with a mass ratio of roughly 2:1 at a lookback time of about 3.5 billion years. The Subhalo ID 476266 also experiences a merger with a mass ratio of approximately 5:1; however, this event takes place at a lookback time of around 7 billion years, which is closer to the estimated time of the GSE major merger episode in the Milky Way.

\begin{figure*}
\centering
\includegraphics[width=1.01\textwidth ]{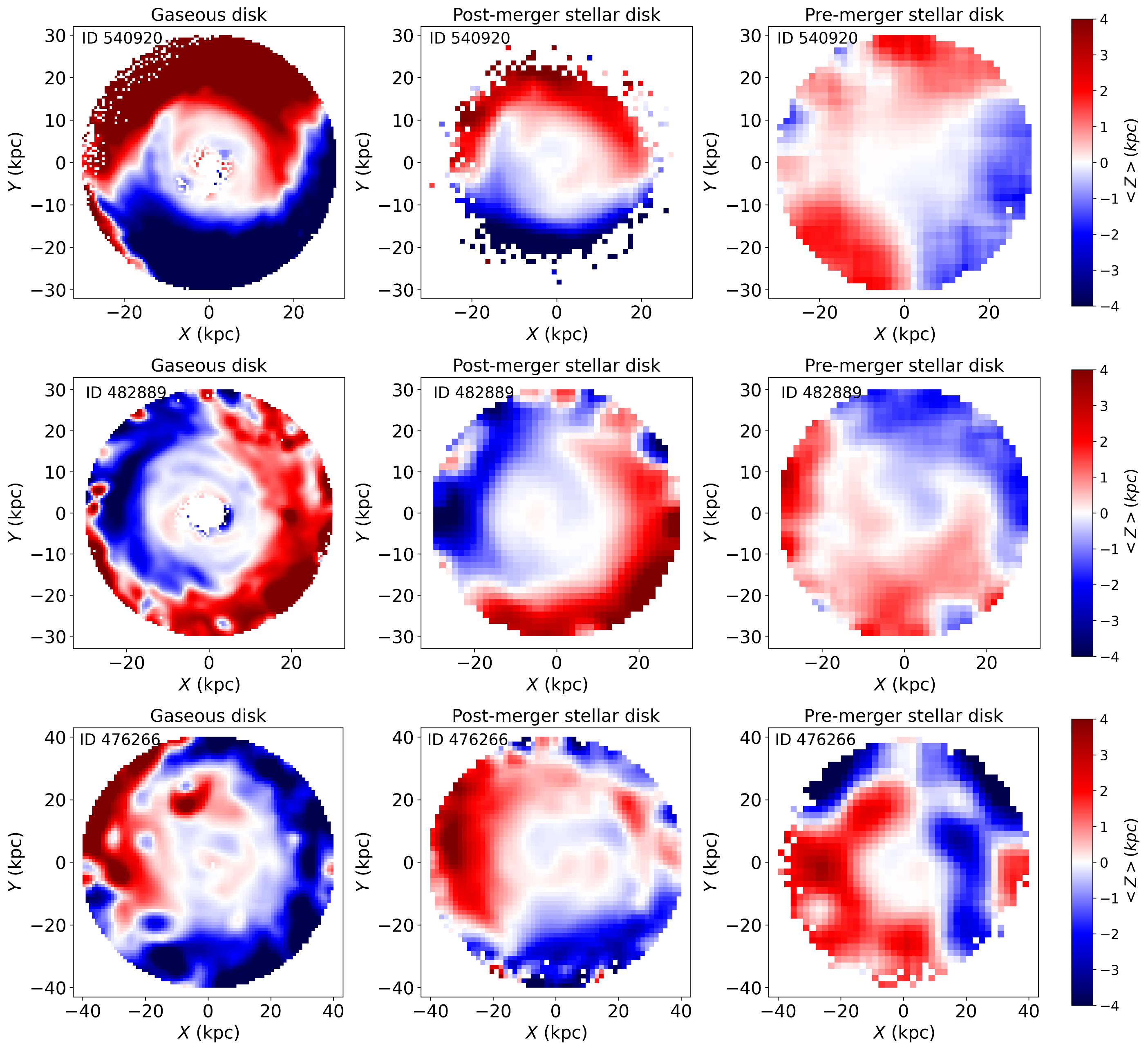}
\caption{X-Y plane maps of average vertical displacement $\langle Z \rangle$ for present-day snapshots (t=0) of Subhalo IDs 540920, 482889, and 476266. These visualizations contrast the warp signatures found within the gaseous disk, stars formed after the merger, and stars formed before the merger. Each map illustrates how the galaxy's vertical structure varies among its distinct stellar populations and gas, shedding light on the influence of the merger event on the observed galactic warp. In each model, the warp in the post-merger stellar disk aligns with the gaseous warp. In contrast, the pre-merger stellar population exhibits a distinct warp signature, appearing at a different orientation.}
\label{fig:fig2}
\end{figure*}

\begin{figure*}[t!]
\centering
\includegraphics[width=1\textwidth ]{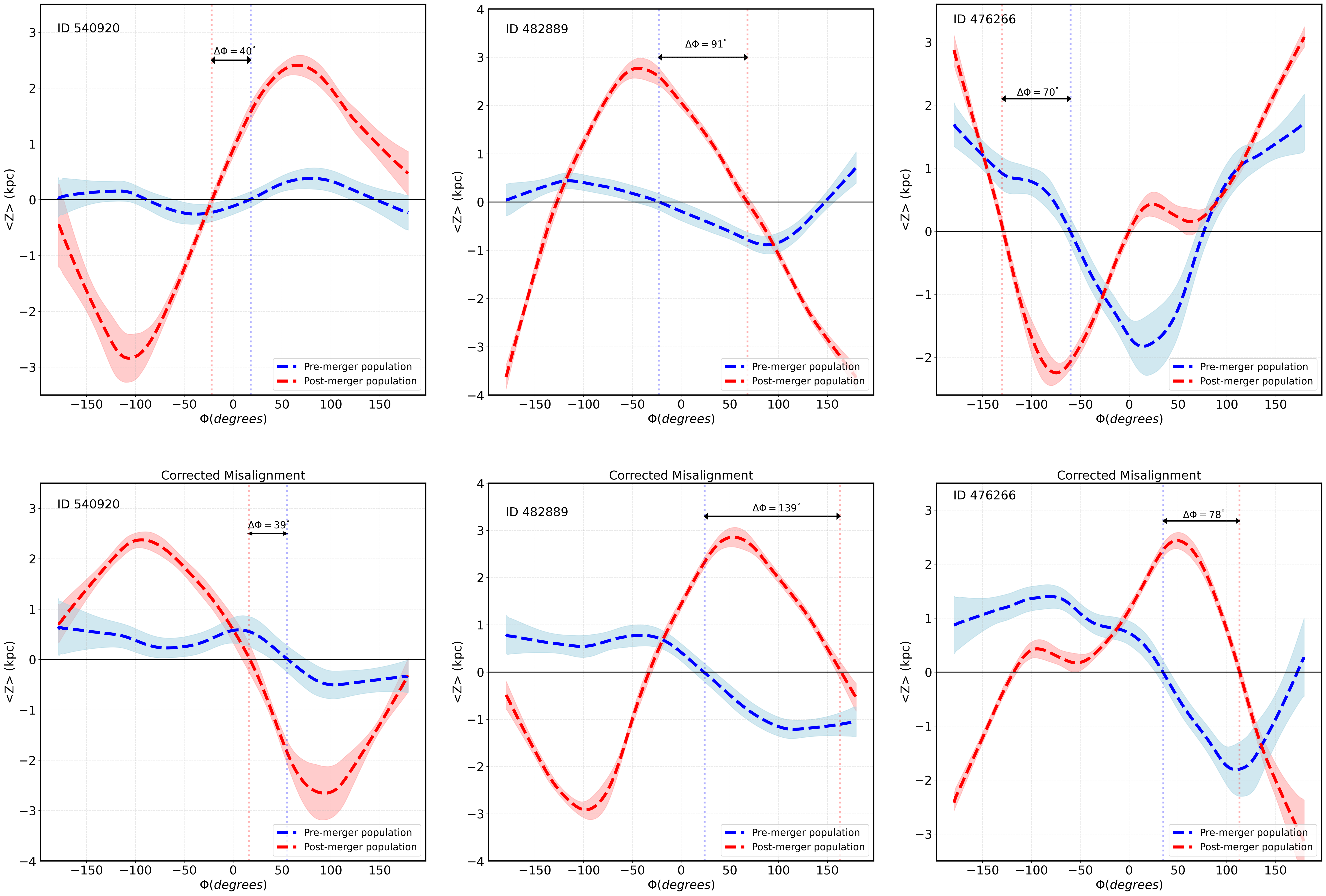}
\caption{Azimuthal variation of the average vertical displacement $\langle Z \rangle$ for the present-day snapshot (t=0) of each model galaxy. \textit{Top panel:} For each galaxy that developed a warp due to a major merger, we analyze the stars contributing to the warp, separating them into populations formed before and after the merger. This plot reveals the differences in warp structure between these two stellar populations, highlighting an intriguing azimuthal misalignment in the warp signature between pre- and post-merger stars. \textit{Bottom panel:} Same as the top panel, except that we corrected for the misalignment due to the differential rotation for pre-merger and post-merger populations.}
\label{fig:fig3}
\end{figure*}

\begin{figure*}
\centering
\includegraphics[width=1.02\textwidth ]{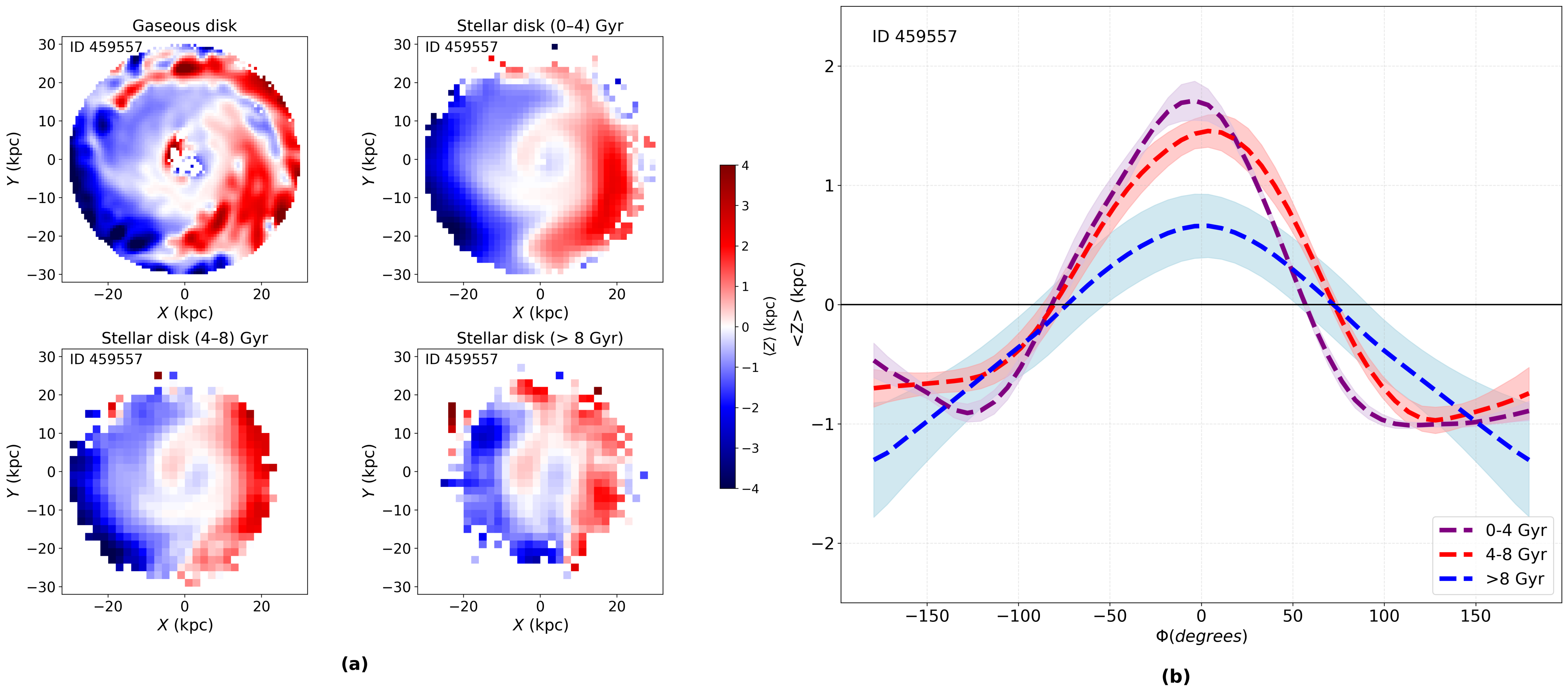}
\caption{Azimuthal variation of the average vertical displacement $\langle Z \rangle$ for the present-day snapshot ($t = 0$) of the model galaxy 459557. As this galaxy does not undergo a significant merger, the stellar population is divided into three age bins 0--4 Gyr, 4--8 Gyr, and $>$8 Gyr to compare their respective warp structures.  (a) Maps of $\langle Z \rangle$ in the X--Y plane for the gaseous disk and stars in each age bin at the present day (b) Azimuthal profiles of $\langle Z \rangle$ for stars in each age bin, illustrating the similarities in the warp signature across stellar populations.}
\label{fig:fig4}
\end{figure*}

\subsection{The misaligned warp}

Figure \ref{fig:fig2} illustrates the X-Y plane colored by the average Z coordinate, $\langle Z \rangle$, for the gaseous disk, post-merger stellar disk, and pre-merger stellar disk of the warped Milky Way models with IDs 540920, 482889, and 476266. For the gaseous disk, we divided the X-Y plane into 100 bins in each direction, while for the post-merger and pre-merger stellar disks, we used 50 bins. The first two models experienced major merger events at around 4 Gyr and 3 Gyr, respectively, while model 476266 underwent a significant merger episode approximately 6 Gyr ago. Consistently, the post-merger stellar disk aligns with the gaseous warp, even when the merger occurred early in the galaxy's evolutionary history. Interestingly, the pre-merger stellar disk exhibits a distinct warp signature with a different azimuthal orientation compared to the post-merger stars, marking precisely the timing of the merger event. It is important to note that $\Delta \Phi$ is only well defined once distinct pre- and post-merger stellar populations exist, i.e. after the merger is complete and the remnant has relaxed into coherent disk structures.

To quantify the misalignment, we look at the average Z coordinate as a function of azimuthal angle $\Phi$. For this, we select the stars in the outer radii where the warp signature begins for each model based on Figure \ref{fig:fig2} and divide the $\Phi$ axis into 200 bins and calculate $\langle Z \rangle$ in every bin. We then applied Locally Weighted Scatterplot Smoothing (LOESS), a non-parametric regression method used to fit a smooth curve through a scatterplot of data points, using the \texttt{statsmodels} implementation in python \citep{statsmodel}. To mitigate the effects of noise and variability, we performed bootstrapping by randomly sampling 200 points with replacement from the set of $\langle Z \rangle$ values computed in each $\Phi$ bin before each LOESS fit. We then applied LOESS with a model-specific smoothing parameter to each bootstrap sample, using five iterations to refine the local fit. This procedure was repeated 200 times, generating an ensemble of smoothed curves from which we quantify variability in the trend. We then computed the mean and standard error of the mean (SEM) at each grid point to effectively convey the central tendency of the data along with its associated uncertainty. 

The results are summarized in Figure \ref{fig:fig3} \textit{(Top panel)}. The blue and red curves show the LOESS mean for pre-merger and post-merger stellar warps. The shading around the mean curve shows the 95\% confidence interval of the LOESS curve calculated by the equation Mean $\pm$ 1.96$\times$SEM. To quantify the misalignment of the warp orientation in the azimuthal direction, we measure the phase shift between the two sine-like curves that represent the warp along $\Phi$ for the two stellar populations. The phase shift, denoted as $\Delta \Phi$, is measured between the two dashed vertical lines, which correspond to the points of intersection of the curves with the horizontal line representing $<Z>$=0 kpc. Note that the phase shift $\Delta \Phi$ is not strictly constant across all azimuthal angles, as the warp profiles are not perfect sine functions. However, this measurement serves as an effective proxy for quantifying the overall offset in warp orientation between the two populations. The plot clearly illustrates two distinct stellar warp structures in galaxies that experienced a major merger event. The pre-merger stellar warp also exhibits a lower amplitude compared to the warp exhibited by the post-merger stellar population. We also  performed a circular autocorrelation analysis on $\langle Z (\Phi)\rangle$ to assess the structure of the warp in post-merger and pre-merger populations for each galaxy. Across all three galaxies, the circular autocorrelation of the post-merger warp profile showed a smooth, symmetric U-shaped pattern, consistent with a coherent $m=1$ warp mode. Whereas the pre-merger warps, though present in all three models, showed more irregular or noisy patterns. For example, ID 540920 lacked symmetry and instead exhibited irregular oscillations, consistent with a more fragmented warp structure, while IDs 482889 and 476266 retained roughly sinusoidal patterns despite added noise compared to their post-merger counterparts. To compute the circular autocorrelation, we applied a standard FFT-based method and normalized the result by the zero-lag value so that the correlation is unity at zero lag. This provides a direct measure of the periodicity and coherence in the warp signal.


Figure \ref{fig:fig4}a shows the X-Y plane colored by $\langle Z \rangle$ for the gaseous disk, along with the warps observed in stellar populations of various age groups: 0-4 Gyr, 4-8 Gyr, and $>$ 8 Gyr for model ID 459557. The analysis shows that the warp signature is consistent across all stellar populations, with no significant azimuthal misalignment between stellar disks of different ages. To further quantify the warp, $\langle Z \rangle$ is calculated as a function of azimuthal angle $\Phi$ for the stellar disk starting at the radius where the warp onset is observed. Figure \ref{fig:fig4}b presents the LOESS curve of mean $\langle Z \rangle$ as a function of azimuthal angle for the different stellar age groups, and the results demonstrate that the warps have similar structures and are aligned with no substantial offsets. This model has not undergone any significant merger events throughout its history, suggesting that the warp has developed due to factors other than a major merger scenario.

\subsection{Effect of Differential Rotation}
For model galaxies that underwent a major merger, we identified two distinct stellar populations: the pre-merger and post-merger populations. It is well established that satellite interactions can dynamically heat stars in a galactic disk, increasing their vertical dispersion and leading to the formation of a thicker disk \citep{toth1992,quinn1993}. In this context, stars that were already present at the time of a major merger experience this heating and attain a larger vertical extent, forming a geometric thick disk. In contrast, stars that form after the merger, in the absence of further major perturbations, tend to remain closer to the mid-plane, creating a geometric thin disk.

In our models, we observe that the premerger stellar population, corresponding to the thick disk, rotates significantly slower than the postmerger stellar population, corresponding to the thin disk. This kinematic separation is also seen in the Milky Way, where the thin disk exhibits faster rotation than the thick disk \citep{vieira2022}. One of the earliest studies proposing a major ancient merger event in the Milky Way was based on such rotational velocity differences \citep{gilmore2002}.

The observed warp misalignment between the premerger and postmerger populations could be caused by several factors. After the merger, continued mass accretion leads to an increase in the rotational velocity of newly formed stars, as expected from the rotation curve relation $v_{\text{rot}} = \sqrt{GM(<r)/r}$. At the same time, vertical heating of pre-existing stars introduces a significant dependence of azimuthal velocity on vertical height $\langle Z \rangle$ above the mid-plane. Although the standard derivation of rotational velocity from the Poisson equation often assumes $\langle Z \rangle$ equals zero \citep{binney2008}, this assumption may break down in the aftermath of a major merger, where stars are heated and displaced far from the mid-plane. Since stars at larger vertical heights tend to rotate more slowly, this can naturally lead to drastic kinematic differences between the thick and thin disk components.

To test whether the observed azimuthal warp misalignment is a true dynamical signature of the merger or merely a consequence of differential rotation, we adopt a corotating frame for each population. For a given radial range, chosen based on where the warp emerges in each model, we compute the average azimuthal velocity $\langle V_\phi\rangle$ of both pre-merger and post-merger components and derive the angular speed as $\Omega=\langle V_\phi\rangle/R$. We then define a rotation-adjusted azimuthal coordinate as $\Phi^{'}=\Phi-\Omega t$ where $t$ is the time elapsed since the merger event. In this frame, we recompute the difference in warp orientation between the two components by analyzing $\langle Z \rangle$ as a function of the adjusted azimuthal angle $\Phi^{'}$. If the misalignment persists in the adjusted coordinate frame after correcting for the average rotation, this would strongly suggest that the warp misalignment is a true dynamical relic of the merger. On the other hand, if the difference disappears, it would indicate that the initial apparent misalignment was at least partially driven by differential rotation. This could result from secular mass growth or the gradual evolution of the galactic potential rather than directly from the merger itself. 

Note that in performing this differential rotation correction, we have assumed a single average angular speed, $\Omega$, adequately represents each stellar population. This approximation simplifies the correction, but in reality, stellar populations exhibit intrinsic velocity dispersions and radial gradients in angular velocity, which could introduce uncertainties in the estimated azimuthal shift. A more precise treatment would involve correcting with a radially dependent $\Omega(R)$, derived from the rotation curve. However, we adopt a mean value here to isolate the dominant effect and maintain consistency with observational strategies, where full rotation curve information may not be available. Although our method captures the dominant systematic effect, future studies employing more detailed, spatially resolved rotation profiles could further refine the robustness of this approach.

The results are summarized in Figure~\ref{fig:fig3}, bottom panel. We find that the azimuthal misalignment in warp orientation between the premerger and postmerger stellar populations persists even after correcting for their differential rotation. In models 482889 and 476266, the azimuthal offset between the warp peaks increases in the corotating frame, indicating that the observed misalignment is not simply a byproduct of differing rotational velocities. In model 540920, the misalignment angle remains largely unchanged before and after the correction. We note that a slowly tumbling halo would tend to induce a common precession of the line of nodes in both populations; the persistence (and in some models the increase) of the age‑differential phase after the Ω‑correction therefore argues that differential precession in a common potential cannot by itself account for the signal (see Figure 3). These findings support the interpretation that the azimuthal misalignment in warp orientation is a genuine dynamical imprint of the merger event. When such a signal remains after accounting for internal kinematics, it provides a compelling diagnostic of past merger interactions and their long-lasting influence on the disk structure.

\section{Discussion and Conclusion}

Our analysis shows that the azimuthal offset in warp orientation between the pre-merger and post-merger stellar populations remains even after accounting for their differential rotation. This suggests that the observed misalignment is not solely a result of the stars' inherent kinematics, but rather a dynamical signature of the major merger event. Specifically, the warp offset remaining the same or increasing in the corotating frame suggests that differential rotation alone cannot explain the phase shift seen between the two populations.

The most plausible explanation is that the misalignment originates from the dynamical response of the disk to the major merger event itself. During such interactions, the infalling satellite perturbs the host galaxy's gravitational potential in a non-axisymmetric and time-dependent manner \citep{ostriker1989,quinn1992}. These disturbances generate vertical displacements and torques, particularly in the outer disk where stars are more weakly bound \citep{chequers2018}. As a result, structures such as bending waves and warps could form \citep{shen2006,vega,donghia2016,sellwood2022}. The pre-merger stars were already present and experienced the full extent of the perturbation. However, post-merger stars form after the potential begins to settle. These post-merger stars form after the gas has redistributed energy and angular momentum and re-settled into a new equilibrium plane that may not match the pre-existing disk's orientation. This time delay between the merger event and the gas settling down means the post-merger stars are born into a different disk plane, which is shaped by the remnant angular momentum and the redistributed mass from the merger event. Consequently, these two populations now occupy distinct warp planes, with differing phase orientations, even though they are spatially co-located. It is important to distinguish between the geometric manifestation of the misalignment, i.e, the stars occupying distinct warp planes, and the underlying dynamical cause, which is the reorientation of angular momentum imparted by the merger \citep{Rogers2022}. Although the observed offset is measured as a spatial difference, it reflects a lasting dynamic memory of the system’s perturbed state.

Warps in stellar disks are long-lived, especially in the outer disk where dynamical timescales are long. Therefore, the memory of the merger-induced disturbance remains imprinted in the relative orientations of the stellar warps, even several gigayears after the merger. Our findings support the interpretation that the azimuthal misalignment between stellar warps is a dynamical signature of past merger events.


This also provides a potentially useful and independent approach to placing constraints on the timing of a major merger event, particularly by analyzing the age distribution of stars showing azimuthal warp misalignment. However, the precision of such timing is ultimately limited by uncertainties in stellar age determinations, especially for older populations. When corrected for differential rotation, the presence of azimuthal misalignment reflects the dynamical imprint left by the interaction and can serve as an indicator of a past perturbation. Furthermore, this method may help determine the origin of warps in galaxies. Our findings suggest an external merger origin, rather than internal or secular processes, in galaxies where a distinct azimuthal offset is observed between the warps of stellar populations as a function of age.

Several mechanisms other than major mergers can, in principle, produce azimuthal asymmetric warps. (i) Long‑lived torques from a misaligned, non-spherical dark matter halo, including a slowly tumbling triaxial figure, can warp the outer disk \citep{ostriker1989,debattista1999,bailin2003}. In such cases, the torque acts coherently on the pre-existing and newly formed stars (via the reoriented gas layer), so at fixed radius one expects a common, slowly decaying line of nodes across stellar ages, unless the halo torque varies rapidly with time. (ii) Forcing by satellites such as the LMC or Sagittarius can excite bending waves and generate azimuthal asymmetries \citep{Weinberg1995,weinberg2006,bailin2003,laporte2018,poggio2020,semczuk2020}. (iii) Prolonged misaligned gas accretion can also tilt the outer disk and set the warp orientation \citep{lopez2002_a}. All three mechanisms can yield asymmetries; however, they naturally predict either age‑independent phases (i, ii) or a gradual drift of phase with look‑back time (iii), rather than the discrete pre/postmerger phase offset we measure. \cite{gomez2017} analyzed Milky Way–like galaxies in cosmological simulations and showed that the vertical structure of stellar discs can depend on stellar age, with tidal interactions producing perturbations across both young and old populations, while misaligned gas accretion affects mainly the younger stars and cold gas. This distinction supports the interpretation that the age dependence of warp morphology can reveal the underlying formation mechanism. In our merger models, the pre‑merger and post‑merger populations occupy two distinct warp planes whose azimuthal offset survives in a corotating frame (Fig. 3), whereas the control galaxy without a major merger shows no age‑differential phase (Fig. 4).

Recent studies by \cite{Han_2023} and \cite{deng2024} suggest that the GSE merger may have triggered the warp in the outer disk of the Milky Way. Applying the method outlined here to the Milky Way provides an independent approach to investigating this hypothesis. By analyzing the azimuthal structure of the warp in older and younger stellar populations, using age or chemistry as a proxy, one can test for an azimuthal misalignment. If such a misalignment persists after accounting for differential rotation, it would support a merger-driven origin linked to GSE. Thus, this approach could potentially constrain the origin of the Milky Way warp along with the timing of its last significant merger.


Detecting such azimuthal misalignments observationally requires accurate stellar age estimates and precise measurements of vertical displacements across the disk. In the Milky Way, current and upcoming astrometric and spectroscopic surveys such as Gaia \citep{gaia2023}, 4MOST \citep{4most2012,4most2019}, and SDSS-V \citep{sdss2024} provide the necessary phase-space and chemical abundance information to separate stellar populations by age or formation epoch. While direct age dating remains more challenging for external galaxies, deep integral field spectroscopy combined with resolved stellar kinematics could allow similar tests of warp misalignment, especially in edge-on systems. However, a significant challenge remains: identifying an azimuthal offset between distinct warp components may require data that span a wide range of azimuthal angles. If the distinct warp components overlap in azimuth near the solar neighborhood, current surveys may lack the spatial extent needed to resolve the misalignment. This limitation could hinder our ability to identify the offset, even if it exists. 

While azimuthal misalignment between pre-merger and post-merger stellar populations is evident in model galaxies where a major merger induces a warp without subsequent significant interactions, we also find a few cases where galaxies undergo a major merger, yet exhibit a coherent warp across all stellar populations. Interestingly, this tends to happen in galaxies that also experience a significant flyby after the merger. Although establishing whether these later interactions are responsible is beyond the scope of this paper, the pattern suggests that they might play a role in reshaping or erasing earlier misalignments. This raises the possibility that the absence of a measurable azimuthal offset could also be a consequence of the complex dynamical history of a galaxy, rather than just a limitation of current observational capabilities. Understanding why this misalignment disappears will require further investigation.

Another interesting direction for future work is to explore whether the azimuthal misalignment between the pre-merger and post-merger stellar warps contains information about the nature of the merger itself, such as the merger mass ratio, spin alignment (e.g., prograde or retrograde), or the angle of the orbit. It would be useful to examine whether such misalignments occur in a range of merger ratios or depend on specific merger conditions. By doing so, we may be able to predict the misalignment angle between the pre-merger and post-merger stellar disks in the Milky Way, if its warp was induced by a major merger such as the GSE.
\par
\vspace{1em}
\textit{Acknowledgments:} We acknowledge funding support from these grants: HST Cycle 30, HST-AR-17053.004. We thank the anonymous referee for their valuable comments and suggestions.\\
\bibliography{sample631}

\end{document}